# Investigating aqueous mineral interfaces using sum frequency generation spectroscopy

## Authors/Affiliations:

Moritz Zelenka[1,2] and Ellen H. G. Backus[1,2,*]

[1] University of Vienna, Faculty of Chemistry, Institute of Physical Chemistry, Währinger Str. 42, 1090 Vienna, Austria.

[2] University of Vienna, Vienna Doctoral School in Chemistry (DoSChem), Währinger Str. 42, 1090 Vienna, Austria.

[*]Corresponding author

## Abstract

Sum frequency generation spectroscopy can be utilized to address the challenge of studying solid-liquid interfaces under ambient pressure and with macroscopic amounts of liquid, while still retrieving molecular level information. By utilizing a non-linear optical process, this method allows to selectively measure vibrations of interfacial species. Within this chapter sum frequency generation spectroscopy is first briefly introduced and then comprehensively evaluated as an investigative tool for solid-liquid interfaces, with mineral-water interfaces as model system. Additionally, it is explained how the actual probing depth of this method depends on both the experimental setup and the properties of the investigated system, such as surface charges.

## Keywords

Vibrational spectroscopy, sum frequency generation, solid-liquid interfaces, mineral-water interfaces, interfacial water structure

## Key points/objectives box

- Understanding basic concepts of sum frequency generation spectroscopy
- Interpreting sum frequency generation spectra of mineral-water interfaces
- Influence of charge on the water structure and the sum frequency generation spectrum
- Concepts of beyond interface contributions in sum frequency generation spectroscopy
- Effect of the probing geometry on the coherence length

## 1. Introduction

Deciphering events at solid-liquid interfaces is key to understanding processes such as heterogeneous catalysis or mineral dissolution and precipitation. Such studies are experimentally quite demanding and pose various challenges not easily to overcome. On the one hand, it is required to investigate interfaces on a molecular level, often requiring ultrahigh vacuum. On the other hand, it is also of great interest to study solid-liquid interfaces under non-artificial conditions, such as ambient pressure and with macroscopic amounts of liquid at the interface. Furthermore, a suitable probing method should also not perturb or interact with the investigated system. Within this chapter we will introduce vibrational sum frequency generation (SFG) spectroscopy as an interface-specific technique, that meets the above-mentioned criteria. The properties, applications, and limitations of this method are discussed using mineral-water interfaces as model system. Mineral-water interfaces are ubiquitous in nature and are subject to numerous different processes, which can alter both the structure of interfacial water molecules and the mineral surface itself. Examples are photocatalytic reactions dissociating water molecules, mineral particles in the atmosphere influencing cloud formation or the dissolution of minerals in contact with water. Based on existing research, it will be discussed how SFG spectroscopy is applied to study aqueous mineral surfaces and how information on the molecular structure, charge and surface reactions of interfaces is obtained.

## 2. A brief guide to SFG spectroscopy

Vibrational SFG spectroscopy has the aim to specifically probe vibrational modes of molecules within the first few interfacial layers. In this technique, two pulsed laser beams are overlapped spatially and temporally at the interface. One of the beams has a fixed frequency in the visible region ($\omega_{vis}$), while the second beam is a tuneable broadband pulse in the infrared (IR) region ($\omega_{IR}$). In a non-linear optical process light is generated at the sum frequency of the two incident beams, $\omega_{SFG} = \omega_{vis} + \omega_{IR}$ (see Figure 1 a)). The desired sum frequency signal can, however, only be produced in non-centrosymmetric media. For centrosymmetric bulk media, e.g. water, introducing an interface to the system intrinsically breaks the preceding symmetry, giving SFG spectroscopy its inherent interface selectivity. The more the centrosymmetry is broken, the more the SFG signal is increasing. Hence it follows, that the detected SFG intensity is proportional to the net polar orientation and polarization of molecules at the interface.[1] This can be expressed in a more mathematical way using the second order non-linear susceptibility $\chi^{(2)}$. $\chi^{(2)}$ corresponds to a tensor containing molecular hyperpolarizabilities, which describe the non-linear relation between the induced polarization and the incident electric field. The measured SFG intensity, $I_{SFG}$, can be expressed as follows:

$$I_{SFG} \propto |\chi^{(2)}|^2 \, I_{IR} \, I_{vis} \qquad (1)$$

with $I_{IR}$ and $I_{vis}$ being the intensity of the infrared pulse and the visible pulse, respectively.[2] In measurements it is usually compensated for the pulse shape and overlap profile of the IR and visible beams by normalising with a reference. Therefore, in this simplified case the generated SFG should ideally only depend on the nature of $\chi^{(2)}$. As the beam intensities, $I_{IR}$ and $I_{vis}$, are unaffected by the symmetry of the probed system, it can be seen that it is actually $\chi^{(2)}$ that is zero in centrosymmetric environments. Two different contributions of $\chi^{(2)}$ are generally distinguished, the resonant and the non-resonant contributions. The resonant contribution is depending on vibrational modes, i.e. when the incident IR pulse coincides with a vibrational mode of a species then $\chi^{(2)}$ is resonantly enhanced. In order to be SFG active a vibrational mode must be both infrared and Raman active. This resonant enhancement of the signal enables SFG spectroscopy in the first place to acquire information on molecular vibrations. To obtain a vibrational spectrum of interfacial species the detected SFG intensity is then plotted as a function of the IR frequency. The second contribution to $\chi^{(2)}$, the non-resonant part, does not significantly change with IR wavelength. It depends rather solely on the incident beam fields and material properties, such as surface plasmon excitations.[1]

## 2.1 Polarization of the incident light

The outcome of an SFG experiment depends on the polarization of the incident IR and visible beams. In experiments the beams are linearly polarized and usually two different cases are distinguished. The light is referred to as s polarized when the oscillation of the linear polarized beam is perpendicular to the plane of incidence. On the other hand, for p polarized light the oscillations of the linear polarized beam coincide with the plane of incidence. Furthermore, for the sum frequency beam also either the s or p polarized component is selectively detected. Different combinations of polarizations probe different tensor elements of $\chi^{(2)}$. For interfaces which are achiral and rotational symmetric around the z-axis just four combinations of beam polarizations are independent and non-zero, denoted as ssp, ppp, pss and sps. This nomenclature orders the light pulses in decreasing frequency, hence ssp means that the measured sum frequency light is s polarized, the incident visible light is s polarized and the incident IR light is p polarized. The choice of beam polarizations is depending on the probed materials. Generally, for dielectric materials, such as $CaF_2$ or quartz, significant electric fields can be both produced with s and p polarized light, while metallic surfaces, such as gold, tend to have larger electric fields with incident p polarized light.[1]

A broad and comprehensive theoretical introduction to sum frequency generation spectroscopy can be found in Ref. [1], a basic and easy approachable introduction is given in Ref. [3].

## 2.2 Heterodyne-detected SFG spectroscopy

As can be seen in Equation (1), conventional SFG spectroscopy measures the intensity of the SFG signal, which is proportional to the absolute square of $\chi^{(2)}$. This has the main drawback that information on the sign of $\chi^{(2)}$ is lost. However, this sign information is a direct indicator how an interfacial group is aligned with the interface, e.g. points "up" or "down" at the interface. Although carefully fitting SFG intensity spectra can often be used to estimate the relative sign between different resonances, measuring the absolute sign is preferable. To resolve this issue, SFG spectroscopy can be extended to heterodyne-detected SFG spectroscopy, which measures $\chi^{(2)}$ directly. Thus, this technique is capable of determining the required sign information, with the disadvantage of being experimentally more demanding. At this point it has to be mentioned that $\chi^{(2)}$ has a real and imaginary component, which we did not need to consider explicitly previously. When analysing data from heterodyne-detected SFG spectroscopy the imaginary and real contributions of $\chi^{(2)}$ are evaluated separately. The imaginary component, $Im[\chi^{(2)}]$, can be seen as an interfacial equivalent to the bulk absorption.[4] Different vibrational modes can have absorption peaks with different signs in $Im[\chi^{(2)}]$ depending on their nature and orientation regarding the interface. In Section 5 we will see how $Im[\chi^{(2)}]$ is used to differentiate between two possible interfacial species. More detailed information about heterodyne-detected SFG spectroscopy can be found in the literature.[5,6]

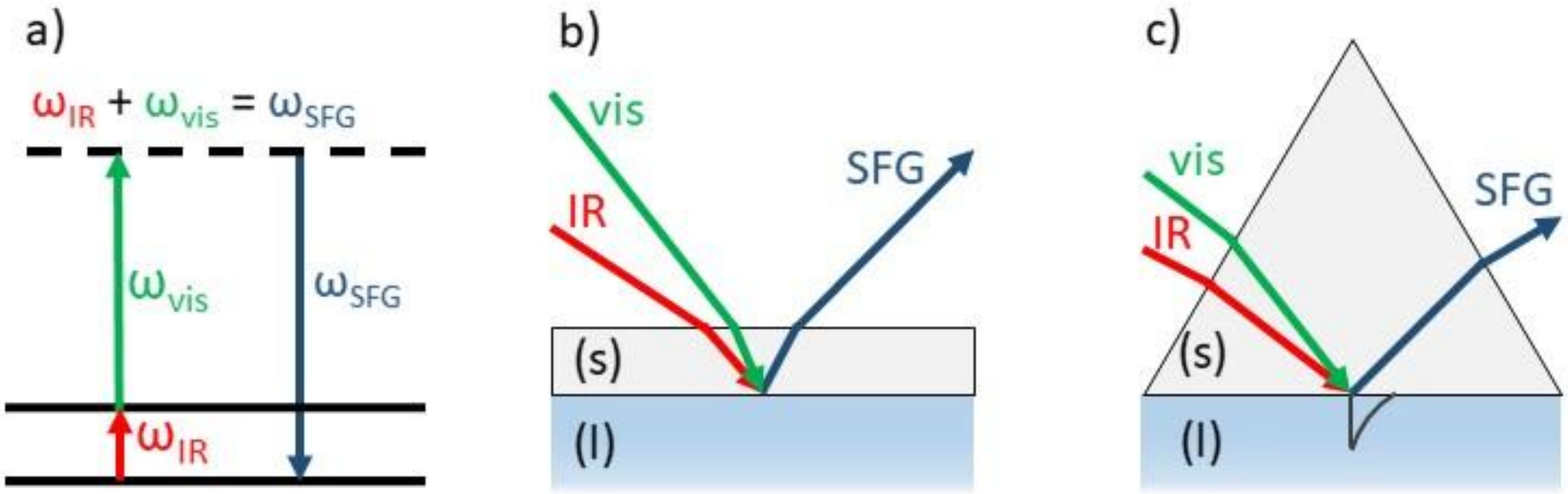

***Figure 1:*** *a) Scheme of sum frequency generation (SFG), where the incident IR beam frequency $\omega_{IR}$ is in resonance with a vibrational excitation. b) Steep angle geometry at the interface of a solid (s) and liquid (l), the reflected SFG beam is detected. c) Evanescent wave geometry at the solid-liquid interface, the reflected SFG is detected.*

## 3. Geometry considerations to measure solid-liquid interfaces with SFG spectroscopy

Before advancing to the interpretation of SFG spectra, some considerations about the measuring geometry are to be made, since the geometry can influence the obtained data. Due to its inherent interface specificity, SFG spectroscopy does not require restrictive artificial conditions, e.g. vacuum or only thin liquid layers. Thus, it is focused here on the geometry of SFG spectroscopy experiments in ambient pressure and with macroscopic amounts of bulk medium. Seemingly trivial, but nonetheless one of the most important considerations is, that the visible and IR beam have to pass either through the solid or liquid medium in order to reach the interface. This should be achieved without extensively attenuating the incident beam intensities. Further, the produced SFG beam should not be absorbed in the medium as well. Many fluids, such as water, absorb significantly in the IR range. Hence, for studying mineral-water interfaces, the IR beam has to penetrate through the solid sample to the interface. The visible beam can then penetrate both through the solid or liquid, usually the experimental setup is simpler if both beams are guided through the solid to the interface. However, solid materials such as minerals, semiconductors or polymers can absorb both visible and IR radiation. If the solid sample is highly absorptive, a thin layer of material can be coated on a transparent crystal, such as $CaF_2$. In this case both the IR and visible beam are passed through the crystal and through a nm to µm thick substrate layer before overlapping at the solid-liquid interface without significant loss of intensity. As the produced SFG beam is both transmitted and reflected at the interface either of the two beams can be detected, depending on the materials used and on the experimental geometry. Two typically used geometries, where the reflected SFG beam is detected, are depicted in Figure 1. In the steep angle geometry (Figure 1 b)) all three beams are below their respective critical angle. On the other hand, in the evanescent wave geometry (Figure 1 c)) one or more beams undergo total internal reflection at the interface.[1,7] The choice of geometry can affect the measured signal and the probing depth of an SFG experiment, which will be discussed in detail in Section 6.

## 4. Vibrational information from mineral-water interfaces

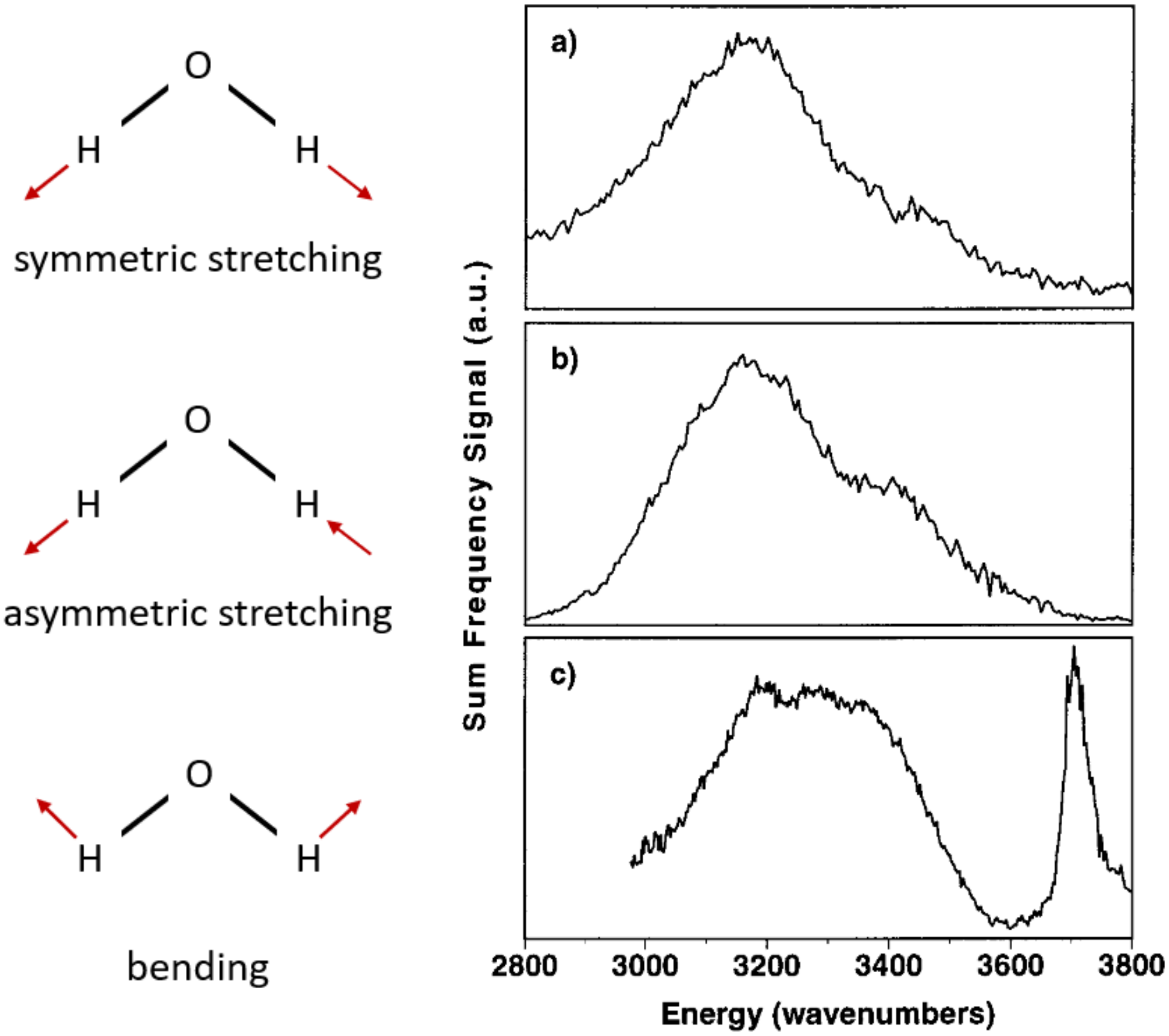

***Figure 2:*** *Left: Vibrational modes of $H_2O$. Right: SFG spectra of water interfaces in ssp polarization. On the bottom axis the frequency of the incident IR pulse is plotted, which coincides with the stretching oscillations of water molecules. a) $CaF_2/H_2O$ interface at pH 3.7, b) $SiO_2/H_2O$ interface at pH 5.1, c) air/$H_2O$ interface. Reprinted with permission from K. A. Becraft and G. L. Richmond, Langmuir 2001 17 (25), 7721-7724, DOI: 10.1021/la011133g. Copyright 2001 American Chemical Society.*

SFG spectroscopy enables molecular insights into interfaces by probing the vibrational modes of interfacial molecules. For mineral-water interfaces these are predominantly water molecules or surface groups of the mineral. In the water molecule, $H_2O$, three vibrational modes are present: symmetric stretching, asymmetric stretching and bending (Figure 2 left). All three are SFG active. Both types of modes, stretching and bending, can be used to obtain information on the water structure and orientation at interfaces. At the mineral surface water molecules can be in distinctively different environments due to varying hydrogen bonding between the molecules and interactions with the mineral surface. This can alter the overall OH bond strengths in the water molecules. As the frequency of a vibrational mode is dependent on the strength of the involved bonds, these intermolecular interactions manifest in changes of the observed vibrational bands for a specific ensemble of water molecules.[8] This shall be rationalized in the following by discussing the stretching modes of water at the $CaF_2/H_2O$ and $SiO_2/H_2O$ interfaces.

## 4.1 O-H stretching modes of water

For the investigation of O-H stretching modes the frequency range between 3000-3700 $cm^{-1}$ is usually of interest. For solid-liquid interfaces the symmetric stretching mode of $H_2O$ is more pronounced than the asymmetric stretching, which often cannot be observed. Figure 2 right a) depicts an SFG spectrum of the $CaF_2/H_2O$ interface at pH 3.7. A broad peak with a maximum in the 3100-3200 $cm^{-1}$ region is observed. This region is associated with water molecules having strong hydrogen bonds with their neighbouring molecules and therefore feature a comparably low vibrational frequency for O-H stretching modes.[9] Upon careful investigation of Figure right 2 a), a second feature, namely a shoulder in the range of 3400-3500 $cm^{-1}$ can be identified. This shoulder is even more pronounced when looking at the $SiO_2/H_2O$ interface shown in Figure 2 right b). The increased excitation energy of these stretching modes is associated with water molecules exhibiting weaker hydrogen bonding. For comparison, Figure 2 right c) depicts a spectrum of the $H_2O$/air interface. This liquid-gas interface again features peaks for water species with different hydrogen bonding environments with maxima in the range of 3100-3400 $cm^{-1}$. In contrast to the solid-liquid interfaces, there is an additional pronounced peak at 3700 $cm^{-1}$ which corresponds to bonds with no to hardly any hydrogen bonding, hence almost gas-like conditions.[10] For the liquid-gas interface this corresponds to OH groups pointing towards the gas phase, which of course cannot be the case for solid-liquid interfaces. However, in several studies weakly coordinated water species, often referred to as free-OH groups, were found at solid-liquid interfaces in the range of 3700 $cm^{-1}$ as well. The presence of such weakly interacting water molecules is also possible at macroscopically hydrophilic surfaces like $SiO_2$. This is exemplified in Figure 3 a) and b), where a spectrum for the $SiO_2/H_2O$ interface is shown without and with preheating to 950°C, respectively. When $SiO_2$ was preheated and then brought into contact with water at neutral pH a peak at 3660 $cm^{-1}$ was found. Supported by theoretical work, this peak was associated to water molecules being close to hydrophobic Si-O-Si pockets, which resulted from heating the sample.[11] However, care has to be taken when assigning such peaks as surface hydroxy groups might be present with very similar vibrational frequencies, also depending on the sample pre-treatment.[4,7,10,11]

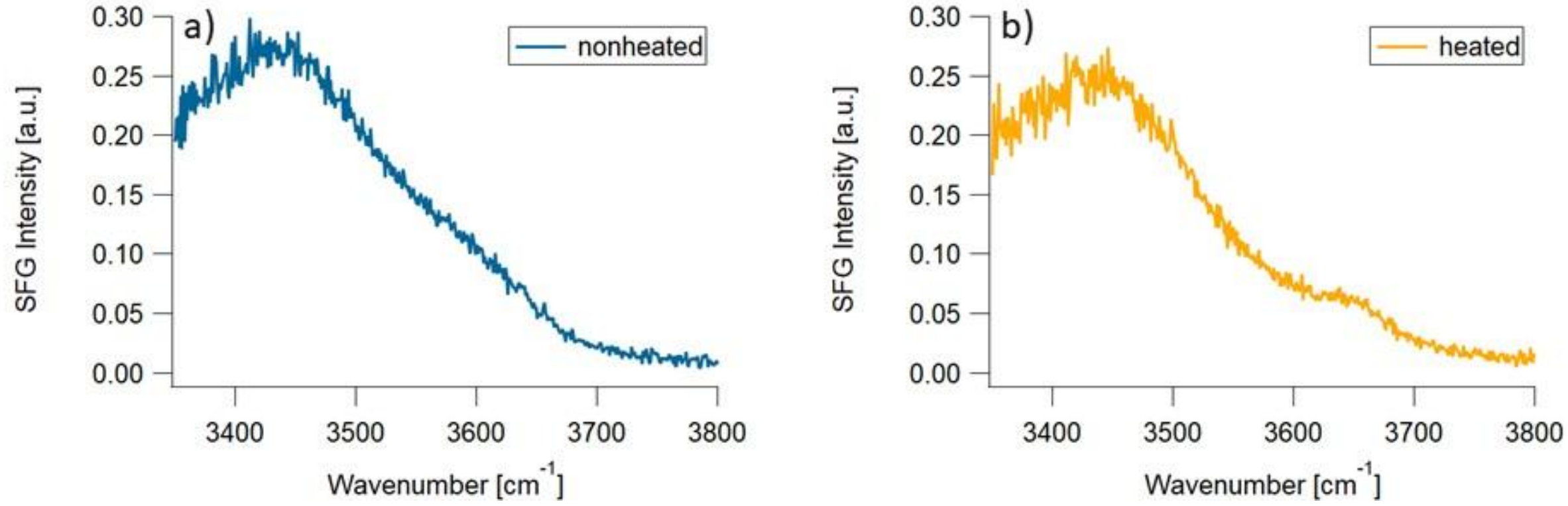

***Figure 3:*** *SFG spectra of the $SiO_2/H_2O$ interface at neutral pH in ssp polarization. The $SiO_2$ substrate was a) not preheated and b) heated to 950°C before the measurement. Data taken from Ref. [11].*

## 4.2 H-O-H bending mode of water

Aside from the O-H stretch mode, the H-O-H bending mode can be utilized to gain information on the molecular interface structure. The interfacial H-O-H bending mode is usually observed in the region of 1550-1700 $cm^{-1}$. Its intensity is weaker compared to the O-H stretching vibration, making it experimentally more demanding to measure. However, studying the water bending mode offers several unique advantages. In more complicated systems, the broad O-H stretching peak can be superimposed by peaks from other species like alcohols, amines/amides or surface bound hydroxy groups. The H-O-H oscillation on the other hand is unique to water molecules and usually well separated from C-O-H, which have lower resonance frequencies. If there is an overlap of vibrational frequencies, with for example C-N-H bending vibrations or amide I vibrations, the modes can be separated by isotopically diluting the liquid with $D_2O$ and measuring the D-O-H vibration instead. Furthermore, the $H_2O$ molecule only has one bending mode, compared to two stretching modes. This means that there cannot be intramolecular bending mode vibrational coupling. Similarly, to the O-H stretching mode the bending mode is also affected by hydrogen bonding. Its frequency is shifted to higher frequencies when there is stronger hydrogen bonding. As the bending mode is less influenced by superposition or coupling with other vibrations, it can be used in models to assess the hydrogen bonding strength more directly than with just the O-H stretching modes.[8,12]

## 5. Effects of charge at the mineral-water interface

Processes such as adsorption, (de)protonation or dissolution often occur when a mineral surface is brought into contact with water. These surface processes are usually more pronounced at either low or high pH values. As a result, the surface can accumulate charge which results in the formation of a surface potential. Different theories exist to model the characteristics of the interfacial region in response to this electrostatic potential. According to the Stern model, the decay of the potential in the liquid is described to be linearly close to the interface and exponentially afterwards in the diffuse layer.[7,13] Such potentials will affect the orientation and/or polarisation of molecules. The $H_2O$ molecule has a permanent dipole moment, as the oxygen atom has a higher electronegativity than the hydrogen atoms. This makes the oxygen partially negatively charged and the hydrogens partially positively charged. Thus, if a mineral surface is charged the $H_2O$ molecules will orient to oppose the outer field. Figure 4 depicts a negatively charged aqueous mineral interface and the two different regions of the decay of the potential. Close to the interface the $H_2O$ molecules are mostly orienting to point with the partial positively charged hydrogen atoms towards the mineral surface. The same effect can be observed for ions in solution, where the positively charged ions are attracted by the negatively charged mineral surface. Additionally, ions will get hydrated by water molecules. As pointed out in Figure 4, ions in solution also screen the surface charge, thus they lead to a faster decay of the surface potential. Hence, two factors can be identified which influence the potential decay: the amount of surface charge, which can be tuned by the solution pH and the ionic strength of the aqueous solution. A measure frequently used to describe screening of charges in electrolyte solutions is the Debye length, as it determines the length until which the potential is decreased to 1/e.

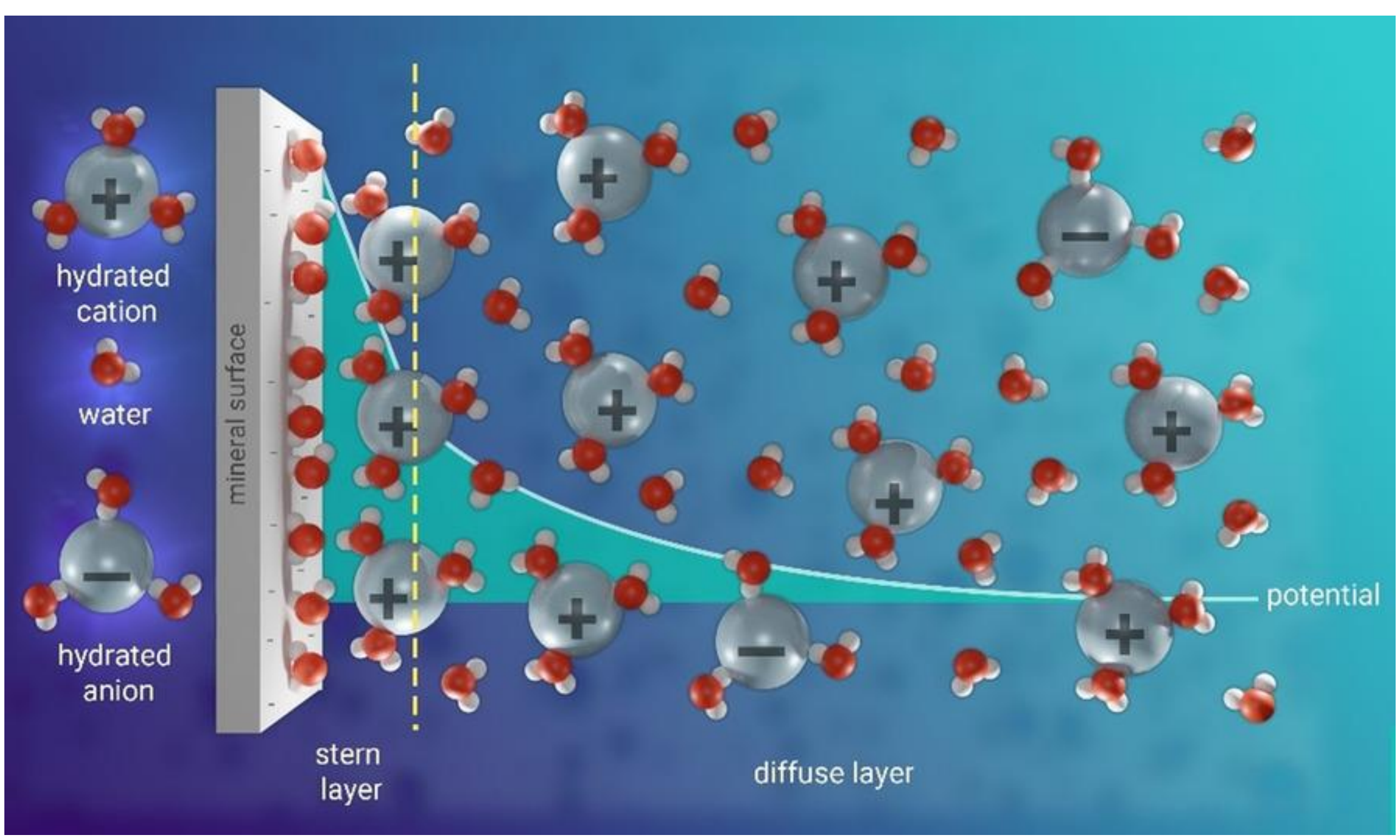

***Figure 4:*** *Model of electrical double layer formation due to a negatively charged surface in contact with an aqueous electrolyte solution. The depth dependence of the potential is plotted according to the stern model, with a linear potential decay in the surface stern layer and exponential decay in the diffuse layer. Reprinted from Ref. [7] under the terms of the Creative Commons Attribution License.*

To exemplify the effects of these interface phenomena on SFG spectroscopy we return to the before mentioned $CaF_2$/$H_2O$ interface. For this discussion we shall remember that the SFG signal is proportional to the degree of net orientation of interfacial molecules. Figure 5 a) shows SFG spectra for three different conditions, acidic (pH 2), neutral (pH 7) and alkaline (pH 13) aqueous solutions. In the acidic case, the $CaF_2$ surface is positively charged and water molecules are more likely to point with their oxygen moieties towards the surface and a significant net orientation of $H_2O$ molecules arises at the interface. Therefore, we also see an intense SFG signal in the measurement. The frequency of these O-H stretching modes matches strongly hydrogen bonded water species. At neutral pH, no O-H stretching peak can be found in the SFG spectrum. Since $CaF_2$ is a hydrophilic interface, this means that there is hardly any net charge at the interface and molecules are stochastically oriented. At pH 13 no peak for strongly hydrogen bonded water species can be observed, as well. However, a less intense peak appears at 3650 $cm^{-1}$. From our knowledge so far, this could be weakly to not hydrogen bonded OH groups from either interfacial water or a surface hydroxy group. With conventional SFG spectroscopy it is difficult to make a decisive assignment in this case. To resolve such problems, heterodyne-detected SFG spectroscopy can be used. $Im[\chi^{(2)}]$ data extracted from heterodyne-detected SFG spectroscopy are shown in Figure 5 b). As previously outlined, the sign of $Im[\chi^{(2)}]$ contains information about the molecular orientation of the band at 3650 $cm^{-1}$, which can be assigned as follows: At pH 2 a broad negative band is observed, belonging to water species collectively oriented with their oxygen atoms towards the surface. At pH 13 the absorption peak at 3650 $cm^{-1}$ has again a negative sign, meaning that the dipole moment of this species is oriented in the same direction relative to the surface normal as in the case of the interfacial water at pH 2. Since the $CaF_2$ interface does not feature a positive charge at pH 13, but is most probably neutral or negatively charged, interfacial water would have no preferential alignment of its dipole moment or the $Im[\chi^{(2)}]$ has a positive sign. On the other hand, a Ca-OH group pointing from the surface towards the bulk would have a dipole moment resulting in a negative sign of the $Im[\chi^{(2)}]$ band, suggesting that this is the species which gives rise to the measured signal. It should be noted at this point, that support from theoretical computations is often essential to allow such conclusions.[4]

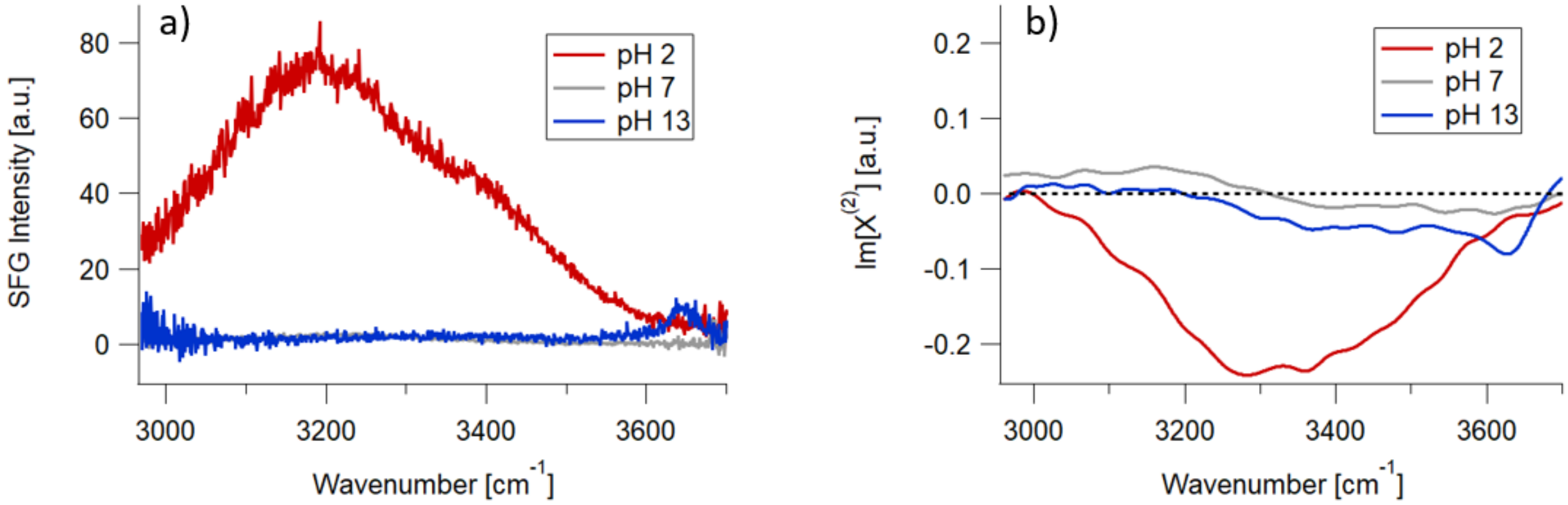

***Figure 5:*** *Measurements of the $CaF_2/H_2O$ interface probed at pH 2, pH 7 and pH 13. In a) SFG intensity spectra are depicted, in b) the respective $Im[\chi^{(2)}]$ component, measured with heterodyne-detected SFG spectroscopy, is shown. Data taken from Ref. [4] under the terms of the Creative Commons CC BY license.*

## 6. Interface and bulk contributions to SFG spectroscopy

The actual probing depth of SFG spectroscopy at the interface is determined by several factors. In the following the effects of charged interfaces, the wavevector mismatch and different probing geometries are discussed. It should be noted that contributions beyond electric dipoles, which make SFG measurements less surface specific, can usually be neglected for liquid phase molecules with strong dipole moments such as water and will therefore not be included in the discussion here. For detailed information about these quadrupole contributions it is referred to additional literature.[14]

### 6.1 Influence of surface charge and ionic strength

As stated previously, measured SFG intensities are proportional to the net polar orientation/alignment of molecules (i.e. a polarization of the medium). In the case of strongly charged interfaces, these effects extend further into the bulk than just the first few interfacial layers. This means that symmetry breaking, necessary for SFG, is not restricted solely to the interfacial region anymore.[15] To account for the resulting expansion of the experimental probing depth, additionally to the 2nd order non-linear response $\chi^{(2)}$, a 3rd order non-linear response term $\chi^{(3)}$ is introduced to describe the measured signal. The 3rd order response depends on the electric fields of the incident visible ($E_{vis}$) and IR beam ($E_{IR}$) and on the electric field induced by the surface potential ($E_0$). Therefore, the electric field of the SFG beam ($E_{SFG}$) is expressed as follows:

$$E_{SFG} \propto \chi^{(2)} E_{vis} E_{IR} + \chi^{(3)} E_{vis} E_{IR} \int_0^{\infty} E_0 dz \,, \qquad (2)$$

assuming for simplicity that $\chi^{(3)}$ is constant along the probed range and that the z-coordinate is perpendicular to the interfacial plane, hence the z-axis equals the probing depth.[16] As depicted in Figure 4 $E_0$ approaches zero at z=∞, due to screening effects by the solvent and ions. With this wider spatial range where SFG photons can be produced another effect has to be considered in our description, namely a deviation of the phase matching condition necessary for SFG. That is the wavevector mismatch, also referred to as phase mismatch, of the IR, visible and SFG photons. The resulting wavevector mismatch $\Delta k_z$ along the probing depth (z-axis) is expressed using the wavevectors of the IR ($k_{IR,z}$), the visible ($k_{vis,z}$) and the sum frequency photons ($k_{SF,z}$):

$$\Delta k_z = |\, k_{IR,z} + k_{vis,z} - k_{SF,z}|. \qquad (3)$$

This relation is also depicted in Figure 6 left in the complex plane for different experimental conditions. For transparent media the three wavevectors are all real, as shown Figure 6 left a) and the magnitude of the mismatch is largest. In the case of aqueous mineral interfaces, i.e. with water as the liquid medium, the IR beam is absorbed and $k_{IR,z}$ gets a nonzero imaginary component, which is depicted in Figure 6 left b). To incorporate the phase mismatch in equation (2), the third order contribution is expanded with a phase factor $e^{i\Delta kz}$ as follows:[17]

$$E_{SFG} \propto \chi^{(2)} E_{vis} E_{IR} + \chi^{(3)} E_{vis} E_{IR} \int_0^\infty E_0 e^{i\Delta k_z z}\, dz\,. \qquad (4)$$

With this equation at hand, we can further discuss the non-interface contributions to SFG spectroscopy and how they are influenced by the composition of the liquid phase. A limit for SFG generation is set either by the symmetry breaking due to charge at the interface expressed by the Debye length or the phase mismatch of the involved pulses indicated by the coherence length. The coherence length $l_c$ depending on the wavevector mismatch is then described as:

$$l_c = \frac{2\pi}{\Delta k_z}\,. \qquad (5)$$

If the Debye length exceeds the coherence length destructive interference can occur between sufficiently spatially separated SFG active molecules, which will diminish the non-interface SFG contribution.[18] Figure 6 right illustrates the interplay between the Debye length and interference effects for a negatively charged silica-water interface. The SFG signal, separated in interfacial and diffuse layer contribution, is plotted against the Debye length and the corresponding ionic strength of the aqueous solution. The ratio of interfacial to diffuse layer contribution is maximized in two areas:

1. Debye length < 1 nm: Due to effective screening of the surface charge, the breaking of symmetry is restricted to the first few interfacial layers and no SFG can be generated in the diffuse layer.
2. Debye length > 100 nm: Destructive interference of SFG generated in the diffuse layer.

This means that in the limit of very low (< 1 µM) and very high (~ 1 M) ionic concentrations it is possible to compensate for effects from electrical double layer formation at charged interfaces and the highest interface specificity is achieved.[14]

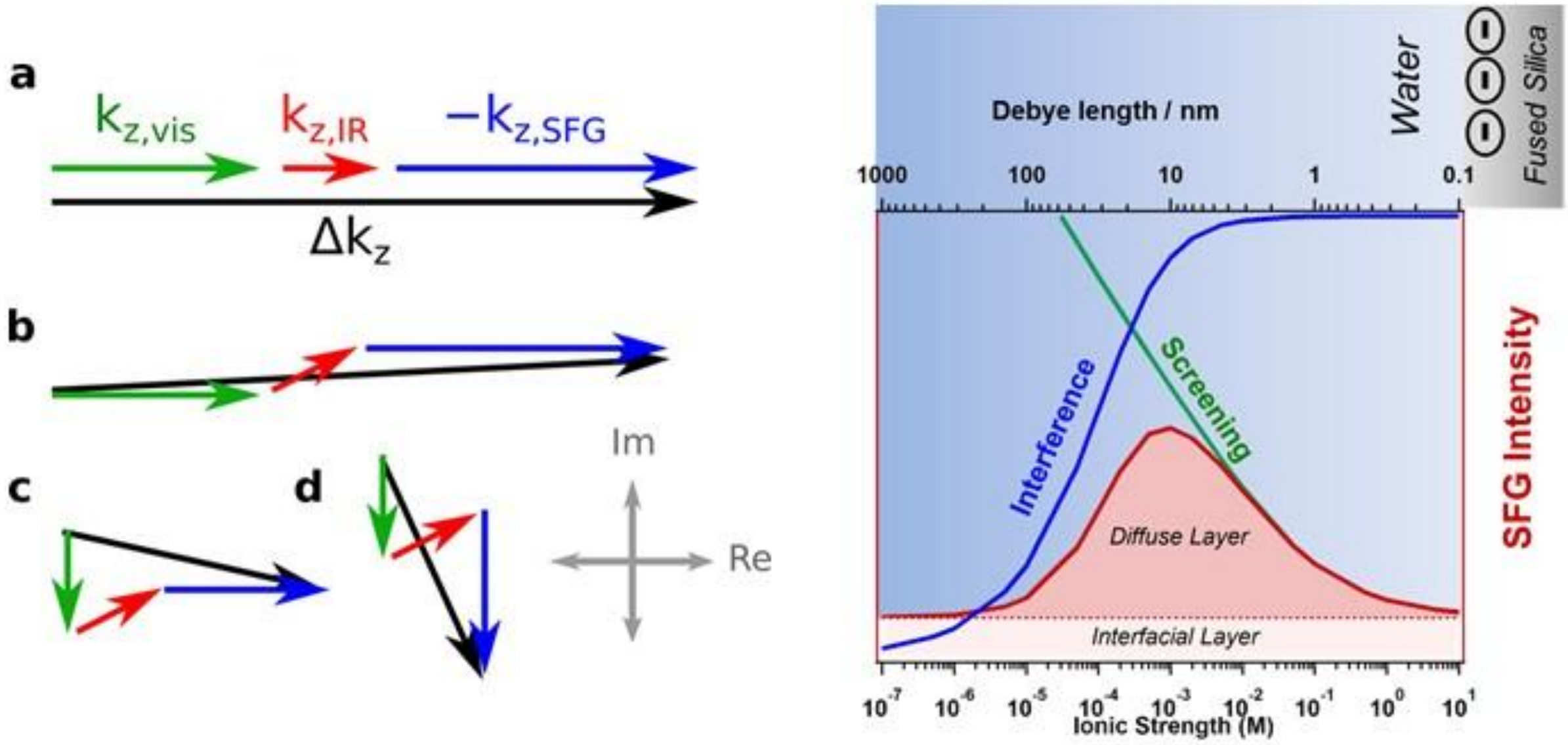

***Figure 6:*** *Left: Scheme of wavevector mismatch of the IR, visible and SFG photons in the complex plane along the surface normal. The following conditions are depicted: (a) transparent medium, with all three beams below their respective critical angels, (b) medium where the IR beam is absorbed, with all three beams below the critical angle, (c) IR absorption and the visible beam is above its critical angle, (d) IR absorption and the visible and SFG beams are above their respective critical angle. Reprinted from Ref.* [17]*. Right: Theoretical model for the SFG intensity simulated at a negatively charged fused silica-water interface. The SFG intensity is divided into the constant contribution from the interfacial layer and the part from the diffuse layer that varies with Debye length/ionic strength because of interference and screening effects. Adapted from Ref.* [14]*.*

## 6.2 Influence of the probing geometry

Another factor influencing the probing depth and contributing to the measured signal is the employed probing geometry. The experimental geometry will not alter the Debye length, but it can very well affect the coherence length depending on the incident angles employed for the visible and IR pulses. This can be rationalised by the magnitude of the phase mismatch, as shown Figure 6 left c) and d). We again assume that the IR pulse is absorbed by water and thus $k_{IR,z}$ inhibits a complex part. Further, we have to consider that above the critical angle total internal reflection occurs, which causes the respective wavevector to shrink and rotate by 90° in the complex plane, due to the exponential decay of the field.[17] Figure 6 left c) shows the case where the angle of incidence for the visible pulse is above the critical angle and Figure 6 left d) where both the visible and SFG pulse angles are above their respective critical angle. It is evident that the wavevector mismatch is smaller in case of geometries involving one or two beams above their respective critical angles, which according to Equation (5) means that the coherence length gets longer.[17] Another factor influencing measurements in the evanescent wave geometry (Figure 1 c)) is the aforementioned exponential decay of the penetrating fields, which can also act as the probing depth limiting factor.[7] Concluding, several factors about the chosen geometry and experimental parameters attribute to the measured SFG signal and for different applications or research questions careful consideration of these factors allow for better insights into the probed system.

## 7. Effects of static and flowing liquids at mineral-water interfaces

A major advantage of SFG spectroscopy is its applicability to a wide range of probing systems and its non-invasive character, compared to experimental techniques that interact with the system, such as atomic force microscopy. Additionally, the liquid phase does not have to be static but can also be flowing. Understanding effects of flowing liquids and comparing the results to static conditions is of interest for a broad range of research fields. This includes environmental research questions, as there is an abundance of flowing water in rivers and oceans, electrokinetic measurements or heterogenous catalysis, where flowing conditions may alter the properties of the catalyst. By comparing static and flowing liquid conditions insights into dissolution and diffusion dynamics can be obtained.[19–22]

As an example, we will once more look at the $CaF_2/H_2O$ interface as depicted in Figure 7 a). SFG measurements with and without liquid flow are depicted in Figure 7 b) for a pH 3 solution. A broad intense OH-stretching peak in the hydrogen bonded region can be observed both when the mineral is in contact with flowing and static liquid, but in the case of liquid flow the intensity of the peak is significantly increased.[20] To explain this increase of SFG intensity we first assume that the net number of molecules at the interface does not change, hence the higher SFG intensity has to come from an increasing degree of molecular alignment at the interface. Since water acts as solvent, it is expected that predominantly water molecules are present at the interface, which in turn have dipole moments. Therefore, the increase of molecular alignment with flowing liquid can be explained by an increase of surface charge upon flow. To shed light into this process, we first look at the case of static liquid. The $CaF_2$ surface (indicated as $\equiv \mathrm{CaF_2}_{(\mathrm{s})}$) can accumulate charge at low pH values due to surface dissolution processes:

$$\equiv \mathrm{CaF_2}_{(\mathrm{s})} \rightleftharpoons \equiv \mathrm{CaF}_{(\mathrm{s})}{}^{+} + \mathrm{F}_{(\mathrm{l})}{}^{-}$$

In the reaction one $\mathrm{F}^-$ ion is dissolved and a positive charge remains at the surface. We can also see that the surface charge is not associated with protonation of the surface at low pH values but rather due to the fact that $\mathrm{F}^-$ dissolution is favoured over $\mathrm{Ca}^{2+}$ due energetically lower hydration of the $\mathrm{F}^-$ ion. When liquid flow is then initiated interfacial liquid is replaced by a fresh solution, which essentially removes the $\mathrm{F}^-$ ions from the chemical equilibrium and the reaction path producing $\mathrm{F}^-$ ions is facilitated.[19] By measuring the SFG intensity as function of time kinetic insights into these surface processes can be obtained as depicted in Figure 7 c). A more comprehensive explanation, which is beyond the scope of this chapter, explaining the underlying phenomena and processes can be found in the literature.[19,20]

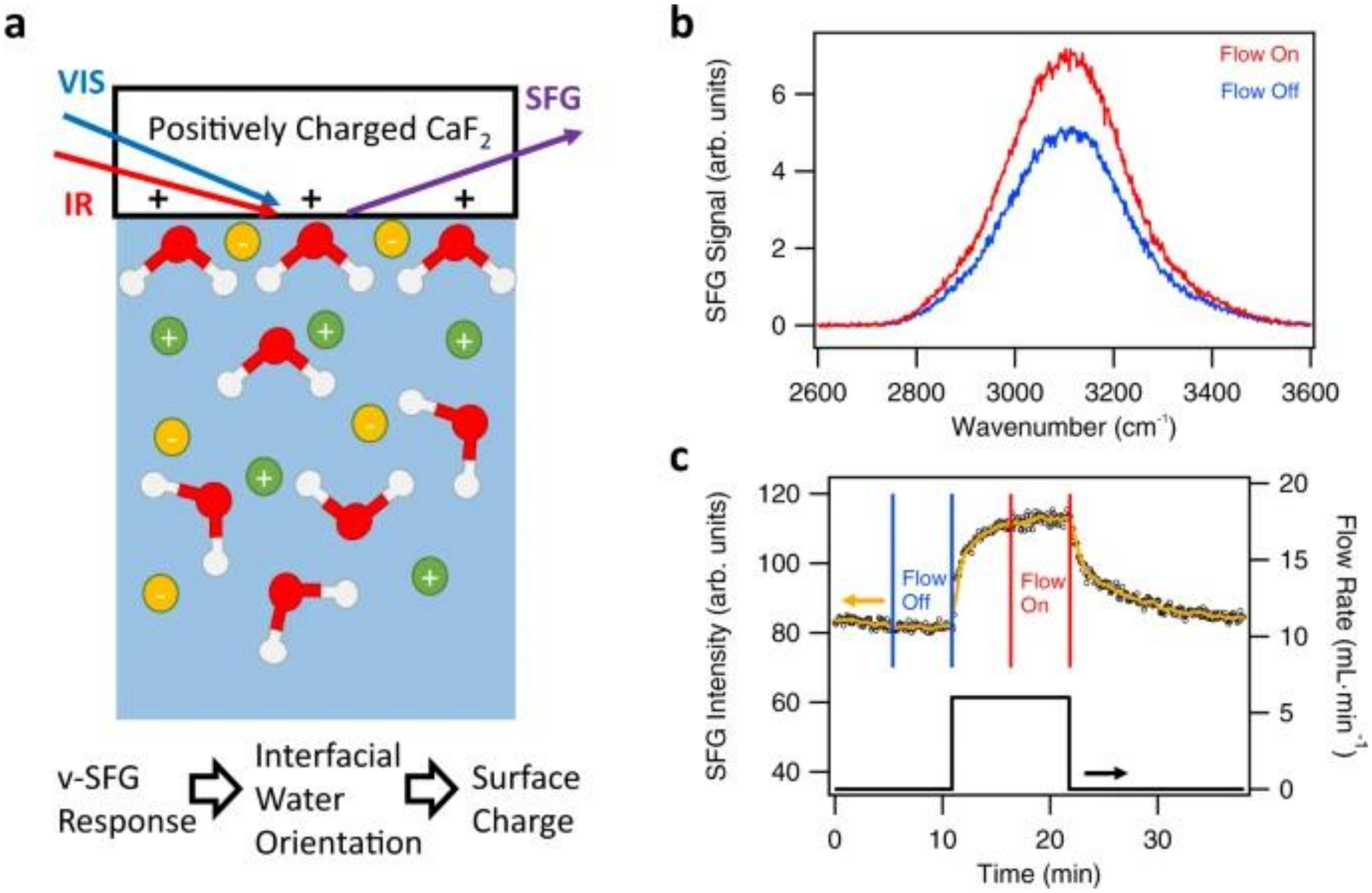

***Figure 7:*** *a) Sketch of the measurement setup for the $CaF_2/H_2O$ interface. The aqueous solution was set to pH 3 using HCl and SFG spectroscopy was done in the evanescent wave geometry using ssp polarization. b) SFG spectra of the OH stretch region recorded with liquid flow in red and with static liquid in blue. c) Integrated SFG intensity as a function of time for one flow on-off cycle, one black dot represents one integrated spectrum. The flow-rate is depicted below. The red and blue vertical lines indicate the steady-state regime for flow-on and flow-off conditions. Reprinted from Ref. [20] under Creative Commons Attribution 4.0 International License.*

## 8. Conclusion and outlook

SFG spectroscopy is a versatile technique for studying interfaces. It is relying on a non-linear optical process that is only allowed when centrosymmetry is broken, i.e. in the case of an interface. Due to this inherent interface selectivity, probing solid-liquid systems experimentally does not require stringent artificial conditions, but can be performed at ambient pressure and with macroscopic amounts of liquid.

Applying SFG spectroscopy to mineral-water interfaces allows for fundamental insights into the water structure, surface groups and dynamic processes like surface dissolution. All three vibrations of the water molecule are SFG active and can be utilized to obtain information on the presence, the hydrogen bonding strength and the vibrational coupling of interfacial species. Additionally, insights into the surface charge or interactions of solvent molecules with the surface can be acquired.

When interpreting SFG spectra several factors that can influence the measurement of solid-liquid interfaces have to be considered. If surface charge is built up at the interface contributions to the SFG signal of molecules in the liquid phase beyond the first few interfacial layers are possible. These effects are influenced by the Debye length of the surface potential and the coherence length of the used laser pulses. The Debye length can be altered by changing the ionic strength of the solution, while the coherence length is depending on the experimental geometry. As a consequence, both at low (< 1 µM) and high salt (~ 1 M) concentrations contributions from non-interfacial molecules can be minimised.

The advanced understanding of such effects in SFG spectroscopy enables studies in a wider range of solid-liquid systems, such as polymer-liquid interfaces. Moreover, reaction dynamics can be studied with so-called pump-probe experiments. In such experiments, the solid substrate or molecules/surface groups at the probed interface are first excited by a laser pulse, e.g. a visible or ultraviolet pulse to study photocatalysis. After variable time delays a visible and IR pulse combine at the interface to produce the SFG signal, allowing to follow the dynamics at the interface on a molecular scale with sub-picosecond time resolution. This type of experiments is essential for understanding reaction mechanisms and timescales of heterogenous catalysed processes, such as photocatalytic water splitting to produce hydrogen. Therefore, SFG spectroscopy has the potential to play a crucial role in delivering fundamental insights into solid-liquid interfaces, being important both for basic and applied research questions.